# Pinning and hysteresis in the field dependent diameter evolution of skyrmions in Pt/Co/Ir superlattice stacks

K. Zeissler[1*], M. Mruczkiewicz[2], S. Finizio[3], J. Raabe[3], P. M. Shepley[1], A.V. Sadovnikov[4,5], S.A. Nikitov[4,5], K. Fallon[6], S. McFadzean[6], S. McVitie[6], T. A. Moore[1], G. Burnell[1], and C. H. Marrows[1]

[1]*School of Physics and Astronomy, University of Leeds, Leeds LS2 9JT, United Kingdom*

[2]*Institute of Electrical Engineering, Slovak Academy of Sciences, Dúbravská cesta 9, 841 04 Bratislava, Slovak Republic*

[3]*Swiss Light Source, Paul Scherrer Institute, 5232 Villigen, Switzerland*

[4]*Laboratory "Metamaterials", Saratov State University, Saratov 410012, Russia*

[5] *Kotel'nikov Institute of Radioengineering and Electronics, Russian Academy of Sciences, Moscow 125009, Russia*

[6] *School of Physics and Astronomy, University of Glasgow, G12 8QQ, United Kingdom*

ABSTRACT

We have imaged Néel skyrmion bubbles in perpendicularly magnetised polycrystalline multilayers patterned into 1 µm diameter dots, using scanning transmission x-ray microscopy. The skyrmion bubbles can be nucleated by the application of an external magnetic field and are stable at zero field with a diameter of 260 nm. Applying an out of plane field that opposes the magnetisation of the skyrmion bubble core moment applies pressure to the bubble and gradually compresses it to a diameter of approximately 100 nm. On removing the field the skyrmion bubble returns to its original diameter via a hysteretic pathway where most of the expansion occurs in a single abrupt step. This contradicts analytical models of homogeneous materials in which the skyrmion compression and expansion are reversible. Micromagnetic simulations incorporating disorder can explain this behaviour using an effective thickness modulation between 10 nm grains.

INTRODUCTION

It is a truth universally acknowledged, that a thin metallic ferromagnetic/non-magnetic interface in possession of a large spin orbit coupling and thus a strong interfacial Dzyaloshiniskii-Moriya interaction (DMI) is in want of the creation of a magnetic Néel skyrmion[1, 2, 3]. Magnetic skyrmions are chiral spin structures which cannot be continuously deformed into another magnetic configuration, such as the ferromagnetic state [4]. Hence, they are topologically stabilised nanoscale structures [1]. This stability, their small size (as small as a few nm [5]), and their mobility under spin-torques at low spin current densities [6, 7, 8, 9], has generated the current research efforts into their usability in novel magnetic information storage technologies.

In the interfacial DMI systems the stabilised skyrmions are of the Néel type. A Néel skyrmion is characterised by the sense of rotation of the spins forming the boundary between the out of plane skyrmion core and the antiparallel surrounding spins. All the boundary spins rotate in a plane perpendicular to the domain boundary. Mathematically, a skyrmion is identified by a non-zero integer winding number *S* given by [10]

$$S = \frac{1}{4\pi} \int \boldsymbol{m} \cdot \left( \frac{\partial \boldsymbol{m}}{\partial x} \times \frac{\partial \boldsymbol{m}}{\partial y} \right) dx dy, \quad (1)$$

where **m** is a unit vector pointing along the local magnetisation direction. Here we consider skyrmions with $S = -1$.

Recent experimental advances have showcased the potential of interfacial skyrmions at room temperature for applications [6, 8, 9, 11, 12, 13]. These works pushed the greatly promising results of Fe on Ir(111) [3] and PdFe bilayer on Ir(111) [5, 14], which exhibit nanoscale skyrmions at sub-liquid nitrogen temperatures, a huge step towards practical applications. This was achieved by increasing the operational temperature range to room temperature and lowering the necessary stabilisation out of plane field into the mT regime. However, a drawback of these polycrystalline room temperature systems has increasingly become apparent in form of pinning which prevents smooth and reliable skyrmion dynamics [9, 15].

Theoretically, the effect of pinning on skyrmion motion has been studied in detail [16, 17, 18, 19, 20, 21]. The main conclusion found was that the pinning depends on the skyrmion velocity. At low current densities the skyrmion motion is influenced by pinning in such a way that previously pinned skyrmions remain pinned. However, when studying the same pinning site but in combination with an already moving skyrmion it was observed that the skyrmion moves around the pinning site [16, 18, 20]. Furthermore, the Magnus force is expected to help mitigate pinning effects. This is expected to lead to good low current density performance of skyrmions [19]. Defects not only influence the motion, pinning also leads to the occurrence of deformation which was observed experimental [15, 22] and theoretical [16].

In a disorder-free system the diameter of a skyrmion shrinks or expands depending on whether a magnetic field is applied antiparallel or parallel to the skyrmion core. The diameter of atomic scale skyrmions, found in a single, defect free, atomic layer of Pd and Fe on Ir(111), roughly scales with the inverse of the applied field *1/µ$_0$(H - H $_0$)*, where *µ$_0$H$_0$* is an offset field [5]. This is in agreement with numerical calculations [23, 24]. However, in a pristine system, expansion (and contraction) is in theory perfectly reversible due to the absence of any in homogeneities that can give rise to pinning. Micromagnetic simulations have shown hysteresis free expansion and contraction when considering pristine ultrathin magnetic layers [11, 13].

In this paper we are conflating pinning affecting skyrmion translation with pinning affecting skyrmion expansion and contraction. Scanning transmission x-ray microscopy (STXM), utilising x-ray magnetic dichroism (XMCD) contrast, was used to study the magnetic field mediated expansion and compression of skyrmions. It was found that disorder in polycrystalline multilayer systems leads to magnetic pinning. This pinning helps to stabilise skyrmions, however it also leads to a field history dependent hysteresis in the observed skyrmion diameter. Micromagnetic simulation shows that disorder, which induces pinning, can take the form of a spatial modulation in the saturation magnetisation. This represents a spatial thickness fluctuation in real samples. Stochastic pinning is thus a double-edged sword, since the same disorder that leads to a stabilisation of skyrmions also explains the relatively high current density needed to initiate skyrmion motion. Furthermore, the pinning makes quantitative comparison of observed skyrmion spin textures with simulations challenging since the full field history must be taken into account in both cases.

RESULTS AND DISCUSSION

*Magnetic Characterisation of the Samples*

The samples we studied were Pt/Co/Ir multilayers with $N$ repeats, of which a schematic is shown in figure 1(a), grown on x-ray transparent silicon nitride membranes by sputtering (see methods section for detail). These were patterned into 1 μm diameter discs for STXM imaging (figure 1(b)), whilst sheet films were retained for magnetic characterisation. The out-of-plane easy axis was confirmed with room temperature magneto-optical Kerr effect magnetometry (see figure 1 (c)). In-plane superconducting quantum interference device vibrating sample magnetometry (SQUID-VSM) (inset in figure 1(c)) was used to measure the saturation magnetisation $M_S$ = 1.2±0.1 MA/m, the exchange stiffness $A$ = 15±1 pJ/m, and the effective anisotropy field $\mu_0 H_{Keff}$ = 0.90+0.05 T (see figure 1 (d)-(f)). The exchange stiffness was extracted from the temperature dependence of the saturation magnetisation (see supplementary information for details). The saturation magnetisation was calculated assuming a 0.7 nm cobalt layer in contact with a 0.3 nm layer of proximity-magnetised platinum [25] resulting in 1.0 nm of magnetic material separated by 2.5 nm of non-magnetic spacer.

The DMI strength of the $N$ = 10 sample was inferred by measuring $D$ on single polycrystalline trilayers. The interfacial DMI is a property of the heavy metal and ferromagnetic material interface, and as such $D$ should not be affected by an increase of $N$ > 1. Figure 1 (d) to (f) shows that the three other main magnetic properties, which are material and ferromagnet thickness dependent, are not sensitive to the repetition number $N$. In particular, the independence of $D$ on the repetition number $N$ is indirectly inferred by measuring effective anisotropy changes with respect to $N$. The effective anisotropy is an interfacial effect and hence a change in the interface quality as $N$ is increased would result in a change in $K_{eff}$. No such dependence was observed; $K_{eff}$ is seen to be constant as $N$ is increased (see Figure 1 (e)). This shows that the interface is not significantly altered and a change in $D$ is not expected. Two trilayer films were measured (Ta(4.8 nm)/Pt(6.2 nm)/Co(0.5 nm)/Ir(1.5 nm) and Ta(3.7nm)/Pt(4.5 nm)/Co(1.0 nm)/Ir(3.0 nm)). Using asymmetric bubble expansion, a technique very susceptible to local defects [26, 27], $D$ was measured to be 0.6±0.1 mJ/m$^2$ for the 0.5 nm cobalt layer and 0.3±0.1 mJ/m$^2$ for the 1.0 nm layer [method as in [28]]. The expected $1/t$ dependence, where $t$ is the thickness, was observed [29]. This results in an expected $D$ of 0.43 mJ/m$^2$ in a system with 0.7 nm cobalt layers. Brillouin light scattering (BLS) was used to confirm D of the 1.0 nm thick trilayer. $D$ was measured to be 1.03 +0.07 mJ/m$^2$; a factor 3.4 increase with respect to the bubble expansion value. BLS returns an average value over a μm length scale and is therefore less sensitive to local defects. Hence when treating polycrystalline samples we can regard $D$ obtained by asymmetric bubble expansion as a lower limit and results obtained using BLS as an average value [26, 27]. Therefore, a revised average value of $D$ of 1.48 mJ/m$^2$ is expected for the sample with 0.7 nm Co layers that was imaged by STXM.

*STXM Imaging*

The spin textures in the samples were imaged using XMCD-STXM. An external magnetic field was applied perpendicular to the sample plane. The disc was initially saturated at ±60 mT and then the field was reversed in incremental steps until a circular bubble was observed (see figure 2). The field then was decreased to 0 mT. The bubble remained stable and observable. Such a zero field stability was also observed by Pulecio et al [15]. The bubble expansion was imaged as the field was reduced stepwise down to 0 mT. The bubble contraction was subsequently observed as the field was increased again. The images in figure 2(a)-(f) show XMCD contrast, extracted from the STXM images, of the disc at different applied magnetic field. The images in figures 2(a)-(c) were taken as the field was increased from 0 mT towards saturation and the images in figures 2 (d)-(f) were taken as the field was decreased to 0 mT. Images 2 (b) and (f) were taken at the same field (-20 mT) but show strikingly different skyrmion bubble diameters. From this observation alone it is obvious that the behaviour is not the straightforward reversible process predicted by theory. The diameter evolution, extracted from the XMCD images, under an increasing and decreasing field is summarized in figure 2(g) and shows a clear hysteretic behaviour. The application of an increasing field leads to a slow shrinking of the skyrmion bubble above 20.0±0.5 mT. No change in the skymion diameter was observed as the magnetic field was reduced until a critical field of -12.5±2.5 mT, below which the skyrmion expands abruptly. The diameter changes from 270±20 nm at low fields to 130±20 nm under fields of several tens of mT.

A skyrmion shrinks or expands depending on whether a magnetic field is applied antiparallel or parallel to the skyrmion core. The diameter of atomic scale skyrmions, found in a single atomic layer of Pd and Fe on Ir(111), roughly scales with the inverse of the applied field $1/\mu_0(H - H_0)$, where $\mu_0 H_0$ is an offset field [5]. Such epitaxial systems can be assumed to be defect free within the range of the few nm over which the skyrmion occurs. This is in agreement with numerical calculations [23, 24]. Micromagnetic simulations of pristine ultrathin magnetic layers have also shown similar trends [11, 13]. The same approach applied to a more disordered, polycrystalline systems is shown as the dashed lines in figure 2(g) which represent fits of $a+(a/(b \mu_0 H - c))$. The $1/\mu_0 H$ relation can loosely be applied to both curves, at least within certain limited field ranges. The fitting parameters however are wildly different between the two. This comes as no surprise, as disorder changes the local energy landscape that the skyrmion encounters, disrupting the predicted reversibility.

*Micromagnetic Simulations*

We have carried out micromagnetic simulations that reveal that a dependence of the skyrmion diameter on the field history experienced by the disc can only be observed when one introduces disorder. The disorder was introduced in form of a spatial fluctuation in the saturation magnetisation. The disc was subdivided into grains of average lateral size of 10 nm (using a 2D Voronoi tessellation [30]). The use of 10 nm grains is discussed in the supplementary information. Each grain was assigned a slightly different saturation magnetisation $M_S$ drawn randomly from a normal distribution centred about a mean value of 1.181 MA/m with a standard deviation of $\delta M/M=3$ % (see figure 3 (a)). This effectively simulates small thickness variations (roughly 0.1 nm) in the sample (see figure 3 (a)) (experimentally 2% to 4% variations in the saturation magnetisation could be seen for 0.1 nm variations in the thickness [29]).

A large skyrmion was stabilised at 50 mT. The field was decreased until the skyrmion was observed to destabilise (around -40 mT) (see figure 3 (b)). The last stable skyrmion state was then used as the initial state and the field was slowly reduced to 0 mT. A clear hysteretic behaviour of the skyrmion diameter with respect to applied magnetic field was observed. Simulations (figure 3 (b)) as well as experimental data (figure 2 (a-b, d-e) and figure 3 (c-h)) show a deviation from the typical circular skyrmion shape at low field. The images shown in figure 3 (c-h) were taken on six separate discs at 40 mT. Simulations confirm the origin of this behaviour to be disorder.

When comparing simulations of pristine discs and of disordered discs. We found that disorder extends the stability of skyrmions well above and below the narrow saturation magnetisation range of a pristine structure (see figure 3 (i)). In this case, the disc was simulated with a single effective thickness of 7 nm (see supplementary material for a discussion on the validity of this approach). A cell size of $x = 2.0$ nm, $y = 2.0$ nm and $z = 7$ nm was used in order to capture the large skyrmion accurately. Skyrmion stability was observed using a range of mean saturation magnetisation values from 0.83 MA/m to 1.40 MA/m in the disordered disc, but only 1.05 MA/m to 1.21 MA/m in the pristine disc. Point 'i' in figure 3(i) shows stable skyrmions despite an average saturation magnetisation of 0.825 MA/m, representing an absolute magnetisation range spanning 0.78-0.97 MA/m. When evaluating the skyrmion shape using the last stable condition (labelled as point i in figure 3 (i)) one finds that the edge of the skyrmion, the domain wall surrounding the bubble, is localised in areas of high magnetisation (see figure 3 (j) and (k)). The skyrmion edge is a region of changing magnetisation and as such lowers the magnetostatic energy. Thus the skyrmion edge lowers the magnetostatic energy most efficiently when localised in areas of high saturation magnetisation. This is demonstrated by the shift of the normal distribution in figure 3(k) when only considering the grains surrounding the skyrmion edge. This effectively acts as pinning sites in the form of energy wells, whilst grains with low magnetisation act as energy barriers. Therefore, thickness variations throughout the samples affect the pinning distribution, and hence have a direct influence on the skyrmion shape and lead to deformation from the ideal circular shape observed both in simulation and STXM imaging.

However, structural disorder has a direct effect on more than the $M_S$. It is to be expected that other magnetic parameters, such as the effective anisotropy $K_u$ (which largely arises at interfaces and so goes as $1/t$), the DMI strength $D$ (also an interface effect with a similar $1/t$ behaviour), and the exchange stiffness $A$ (reduced through finite size effects at these very low thicknesses), will also be affected by it. We have investigated what these potential effects might be in further micromagnetic simulations. In contrast with non-uniformity in $M_S$ and $D$, where the localisation of skyrmion boundaries pins in regions of higher values, non-uniformity in $K_U$ and $A$ leads to localisation of skyrmion boundaries in regions with lower values, as shown in figure 4. Skyrmion boundaries are regions where the magnetisation direction is changing and hence carry an anisotropy and exchange energy penalty. This leads to pinning of the skyrmion edges in areas of low $K_U$ or $A$ grains where this penalty is least. On the other hand the domain wall energy is lowered in regions of high $D$ (the domain wall energy is given by $4(AK_{eff})^{1/2} – \pi D)$. All four of these disorder models (grain-to-grain variations in $A$, $K_U$, $D$, and $M_S$) lead to qualitatively similar behaviour. This is in contrast to small non-uniformity ($K_U/K=1\%$) which has previously been shown to be non-critical to the skyrmion behaviour [31]. All four are likely to be present in real systems to some extent, and it will be an interesting avenue of future work to determine which is predominant.

CONCLUSION

In conclusion, disorder, which inevitably occurs in polycrystalline samples, aids skyrmion stability. However, disorder has two implications on the optimisation and characterisation of future devices. Firstly, it makes it difficult to evaluate the DMI strength by the method of comparing observed bubble diameters with simulations, since the disorder and the field-history effects that it gives rise to have to be accurately taken into account. This is especially the case for superlattices with large repeat numbers where the roughness can change as the number of layers is increased [32]. Disorder and hence pinning has previously been seen to impede device performance resulting in higher than expected current density needed to drive skyrmion motion. A threshold current density of $2\times10^{11}$ Am$^{-2}$ was observed by Woo et al. [9] in Co and CoFeB interface systems with 15 repeats. Similar high current densities were measured in (Pt/CoFeB/MgO) x15 interface systems [8]. In comparison, single trilayer amorphous materials such as $Co_{20}Fe_{60}B_{20}$ showed threshold current densities in the order of $1\times10^9$ A/m$^2$ [6]. Skyrmions in Pt/Co/Ir systems can manifest with smaller diameters than their counterpart in single trilayer $Co_{20}Fe_{60}B_{20}$ and hence are more desirable [(30-300 nm [13] versus 700-2000 nm [6]]. This shows that achieving a proper understanding of pinning effects is of the utmost importance for the future of skyrmion devices.

METHODS

The thin films were deposited by DC magnetron sputtering in a vacuum system with a base pressure of $2\times10^{-8}$ mbar. An argon gas pressure of 3.2 mbar was used during the sputtering and typical growth rates of around 0.1 nm/s were achieved. The superlattice stack, [Co (0.7 nm)/Ir (0.5 nm)/Pt (2.3 nm)]$_{\times N}$, was grown on a seed layer of 4.6 nm Ta/7.2 nm Pt and capped with 1.3 nm of Ir (a schematic shown in figure 1(a)). The value of $N$ ranged from 1 to 10. The patterned structures were grown on 200 nm thick $Si_3N_4$ membranes, with a thin film simultaneously sputtered onto a thermally oxidized Si substrate (with oxide layer thickness of 100 nm) to provide a witness sample for characterisation of material properties. The layer thicknesses were calibrated using X-ray reflectivity on calibration samples sputtered on $SiO_2$. 1000 nm diameter discs were fabricated using electron beam lithography techniques with a bilayer resist lift-off recipe. The bilayer consisted of diluted copolymer methyl methacrylate (MMA(8.5)MAA) with ethyl lactate (EL11) (4:6 ratio) and a top layer of polymethyl methacrylate (PMMA) 950 A4 was used. Both were spun at 4000 rpm for 40s and baked at 180°C for 5 minutes. The pattern was written with a Raith 50 electron beam lithography tool at 30 keV with a 20 μm aperture in 4 nm steps with an area dosage of 275 μC cm$^{-2}$ and a step size of 6 nm. The disc pattern was developed for 90 s in a 3:7 ratio of deionised water and isopropyl alcohol (IPA) for 90 s and then rinsed with IPA for 30 s. The structures were lifted-off in acetone. Figure 1(b) shows a scanning electron microscopy (SEM) image of a completed disc.

Scanning transmission X-ray microscopy (STXM) at the PolLux (X07DA) beamline at the Swiss Light Source [33] was used to image the out-of-plane magnetic contrast of the nanodiscs. A Fresnel zone plate with an outermost zone of 25 nm was employed to focus the X-rays on the sample, giving a spatial resolution on the order of 30 nm. The images were acquired at room temperature under the influence of a static magnetic field applied perpendicular to the sample surface and parallel to the incident X-rays, which were tuned to the Co $L_3$ absorption edge (ca. 778 eV). Magnetic contrast was achieved employing the x-ray magnetic circular dichroism effect taking the difference between the absorption of left and right circularly polarized X-rays and dividing it by the sum of the absorption images. This leads to a black and white contrast indicating magnetic moments aligned parallel or antiparallel to the incident X-rays. The disc was initially saturated at $\pm60$ mT and then the field was reversed in incremental steps until a circular bubble was observed (see figure 2). The field then was decreased to 0 mT. The bubble remained stable and observable. Zero field stability was also observed by Pulecio et al. [15]. The bubble expansion was imaged as the field was reduced stepwise down to 0 mT. The bubble contraction was subsequently observed as the field was increased again.

The micromagnetic simulation package MuMax$^3$ [30] was used to simulate the experimental results. The simulations were run in the static regime which calculates energetic minima neglecting dynamic effects. The mesh size and magnetic simulation parameters were chosen to be $x = 2$ nm, $y = 2$ nm, $z = 0.7$ nm. Further input parameters were as follows: $M_S$=1.181 MA/m, $K_u$=1.41 MJ/m$^3$, $A$= 14.8 pJ/m and D=1.47 mJ/m$^2$, leading to an exchange length of about 3.3 nm, therefore justifying the discretization grid employed here. The layer thickness was set to 0.7 nm. Periodic boundary conditions along the z axis were used to simulate a finite number of periodic images (macro geometry approach) [34, 35]. Ten 0.7 nm magnetic layers separated by 2.8 nm were simulated.

AVAILABILITY OF DATA
The data associated with this paper are openly available from the University of Leeds data repository, http://doi.org/XX.XX.XX.XX

ACKNOWLEDGEMENTS

Support from European Union grant MAGicSky No. FET-Open-665095.103 and EPSRC grant EP/M000923/1 is gratefully acknowledged. Part of this work was carried out at the PolLux (X07DA) beamline of the Swiss Light Source. Research has been further co-funded by the EU FP7 SASPRO Programme (REA Grant Agreement No. 609427, 1244/02/01) and by the Slovak Academy of Sciences. Part of this work was supported by the Grant from Russian Science Foundation (No. 14-19-00760). A.V.S. acknowledges support from the Scholarship of the President of RF (SP-313.2015.5) and Grant of the President of RF (MK-5837.2016.9). The authors are grateful for C. Moutafis for experimental time.

AUTHOR CONTRIBUTIONS

KZ and PMS fabricated the samples. KZ did the magnetic characterisation measurements. KZ, PMS, SF and JR did the STXM imaging. KZ and GB did the STXM imaging analysis. AVS and SAN did the BLS measurements and analysis. KF, SM$^c$F and SM$^c$V performed the TEM imaging and analysis. MM did the micromagnetic simulations and analysis. The manuscript was written by KZ and MM. All the authors reviewed the manuscript. CHM conceived the experiment with TAM, GB and KZ.

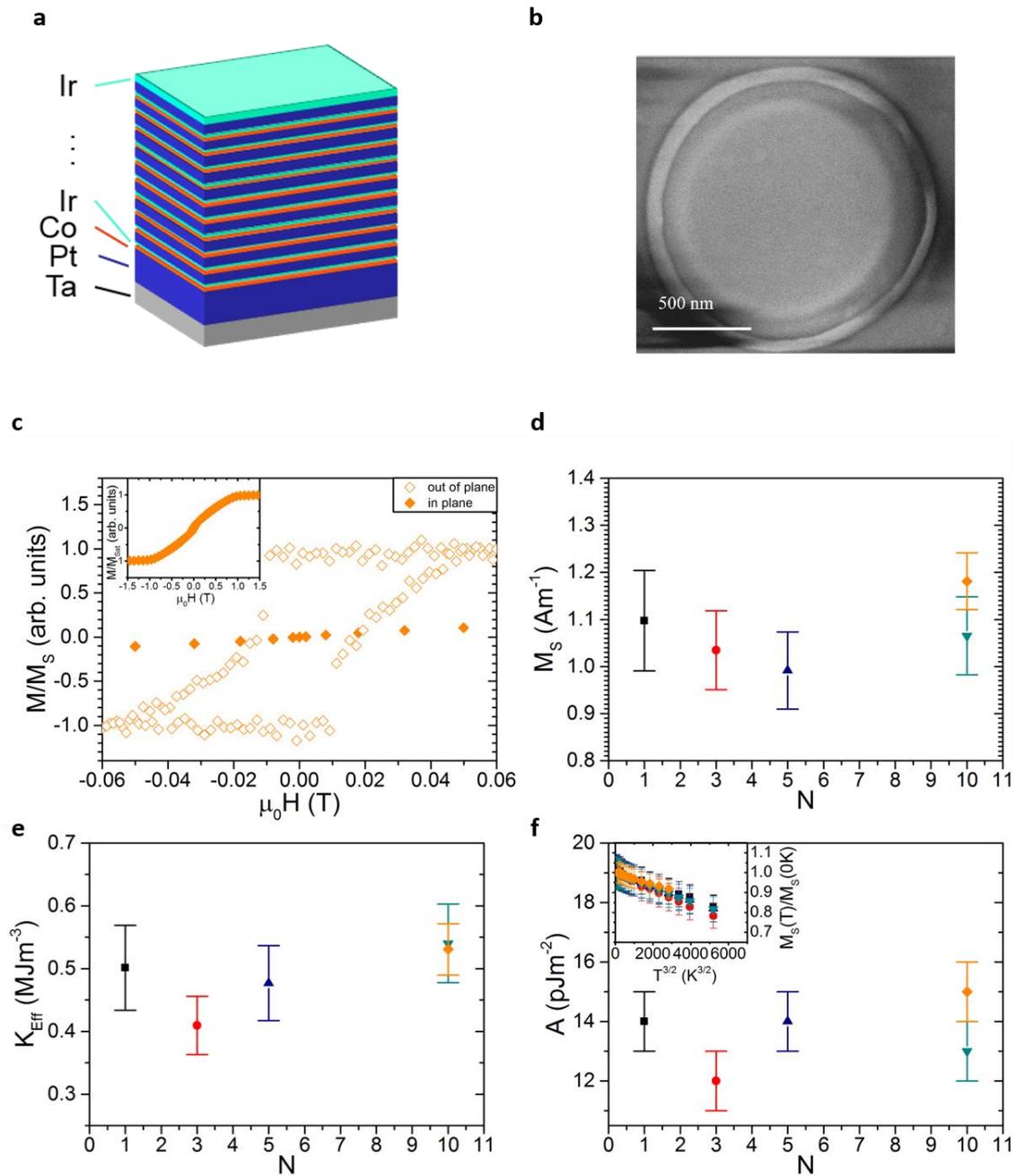

Figure 1: Magnetic properties of [Co (0.7 nm)/Ir (0.5 nm)/Pt (2.3 nm)]$_{\times N}$ stacks. (a) Schematic of multilayer sputtered for $N$ = 10. (b) Scanning electron micrograph of a 1 μm diameter disc patterned from such a multilayer that was subsequently imaged using STXM. (c) Magnetisation versus field showing out-of-plane anisotropy using polar Kerr effect magnetometry and (inset) in-plane SQUID-VSM magnetometry of an unpatterned $N$ = 10 multilayer used to pattern nanodisc. (d) Saturation magnetisation with respect to number of trilayer repeats $N$. (e) Effective anisotropy and (f) exchange constant dependence on repeat number $N$. The inset in (f) shows the linear dependence of the saturation magnetisation (far below from the Curie temperature) on temperature to the power of 3/2.

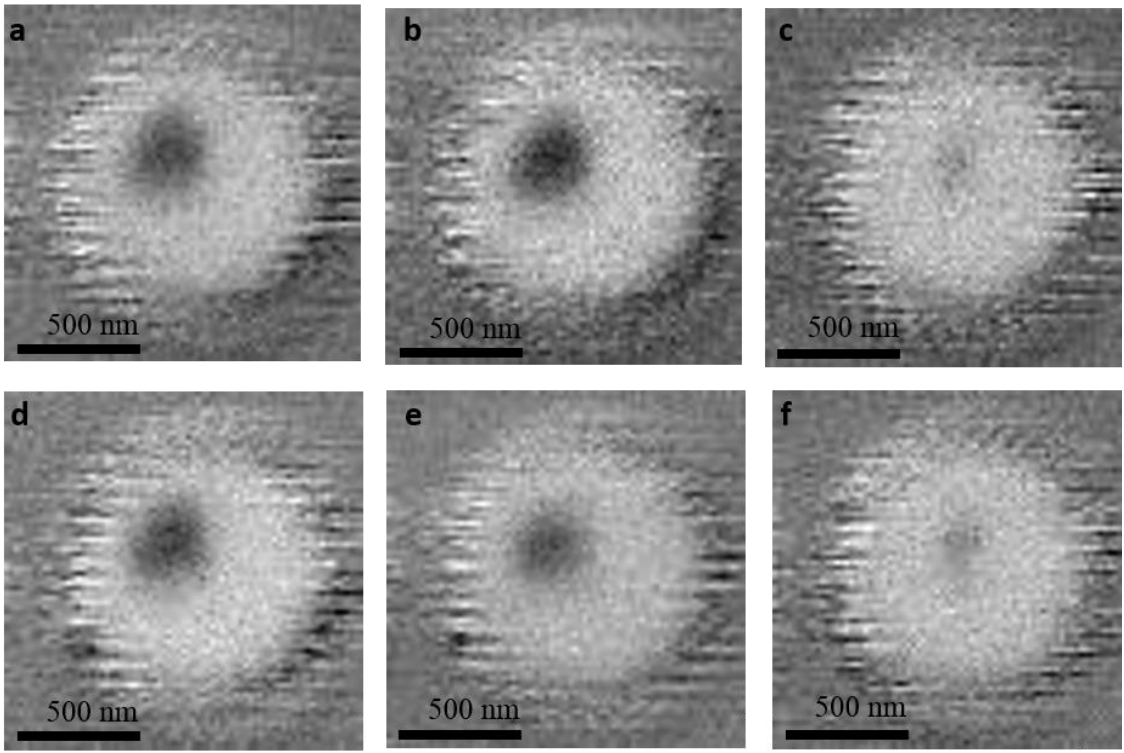

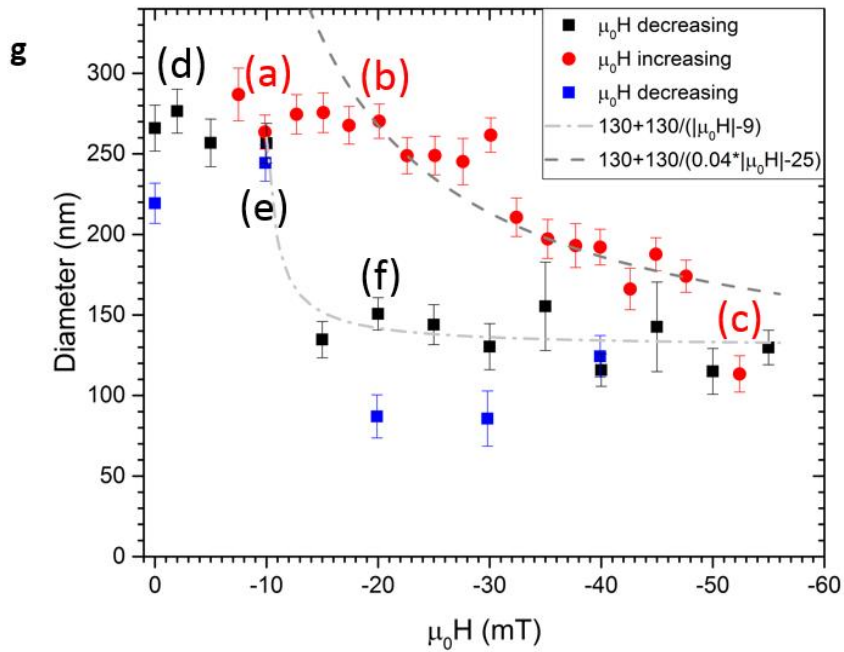

Figure 2: Compression and expansion of a skyrmion in a [Co (0.7 nm)/Ir (0.5 nm)/Pt (2.3 nm)]$_{\times 10}$ multilayer 1000 nm disc imaged using STXM. (a) –(f) show snapshots taken at -10 mT, -20 mT, -52 mT, 0 mT, -20mT and -10 mT, respectively. Light and dark contrast shows antiparallel out-of-plane magnetized domains. (g) Skyrmion diameter versus applied field with respect to sweep direction. A clear hysteresis can be observed. Letters indicate the points in the hysteresis loop at which the images shown in the earlier panels were acquired. The dashed and dashed-dotted lines represent attempts to fit a simple $1/\mu_0(H-H_0)$ law to the data in each branch.

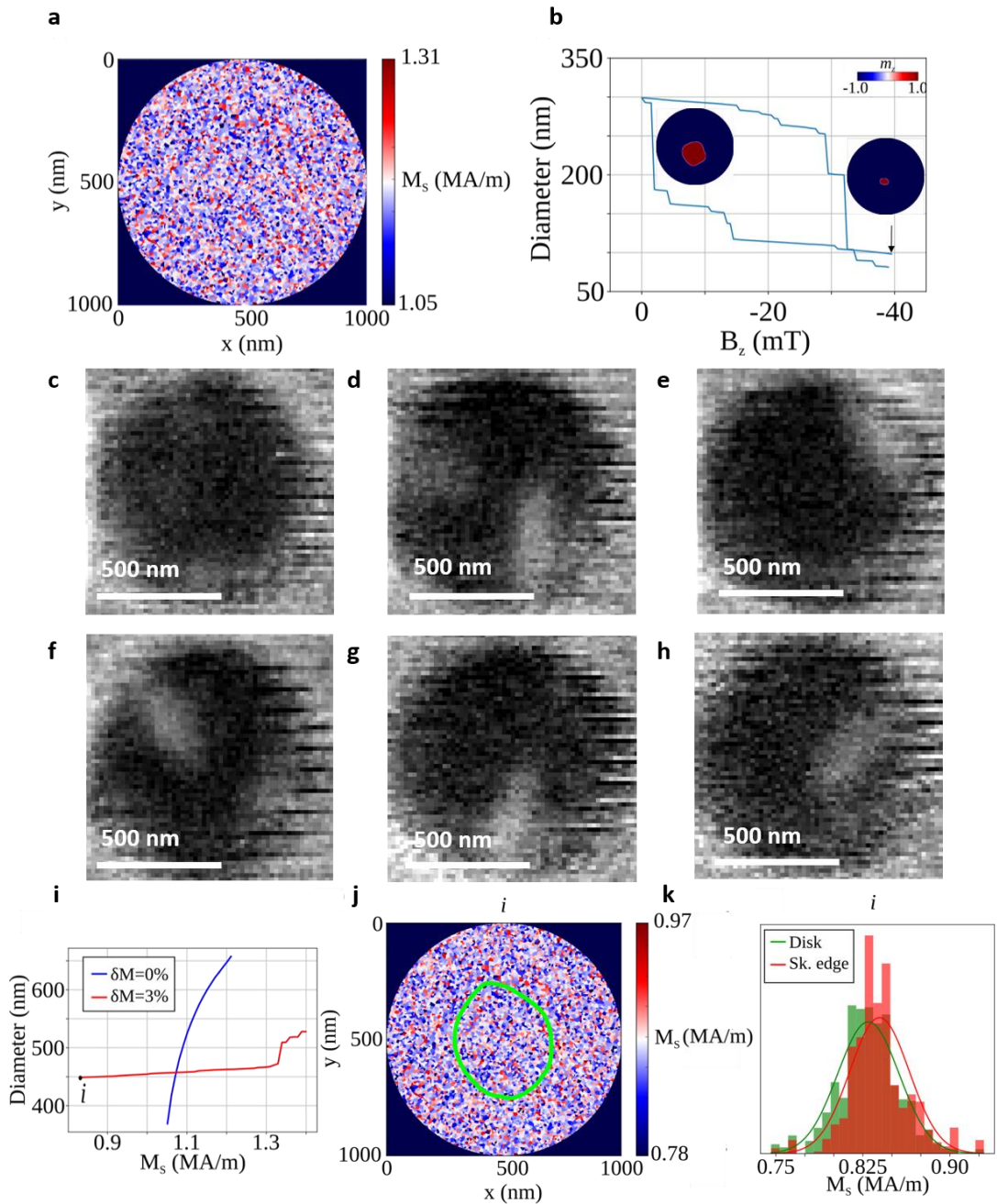

Figure 3: Importance of disorder in magnetic multilayer systems for the formation of skyrmions. (a) Grain structure used in the micromagnetic simulations. (b) Simulated skyrmion diameter as function of magnetic field in an $N$ = 10 multilayer of 0.7 nm Co layers separated by 2.8 nm spacers. The magnetisation was varied by $\delta M$=3% within grains of average size 10 nm. Insets show the simulated skyrmion shapes at low and high field. (c-h) XMCD contrast images of six different 1000 nm discs at 40 mT showing various skyrmion/domain sizes and shapes. (i) Simulated skyrmion diameter dependence with changing saturation magnetisation with and without disorder. (j) Magnetization distribution in the simulation at point "$i$" in (i). The boundary of the stabilised skyrmion is marked by the yellow dotted line. (k) Magnetisation distribution for the structure at point "$i$" considering the whole disc (green) and considering only the skyrmion edge grains (red).

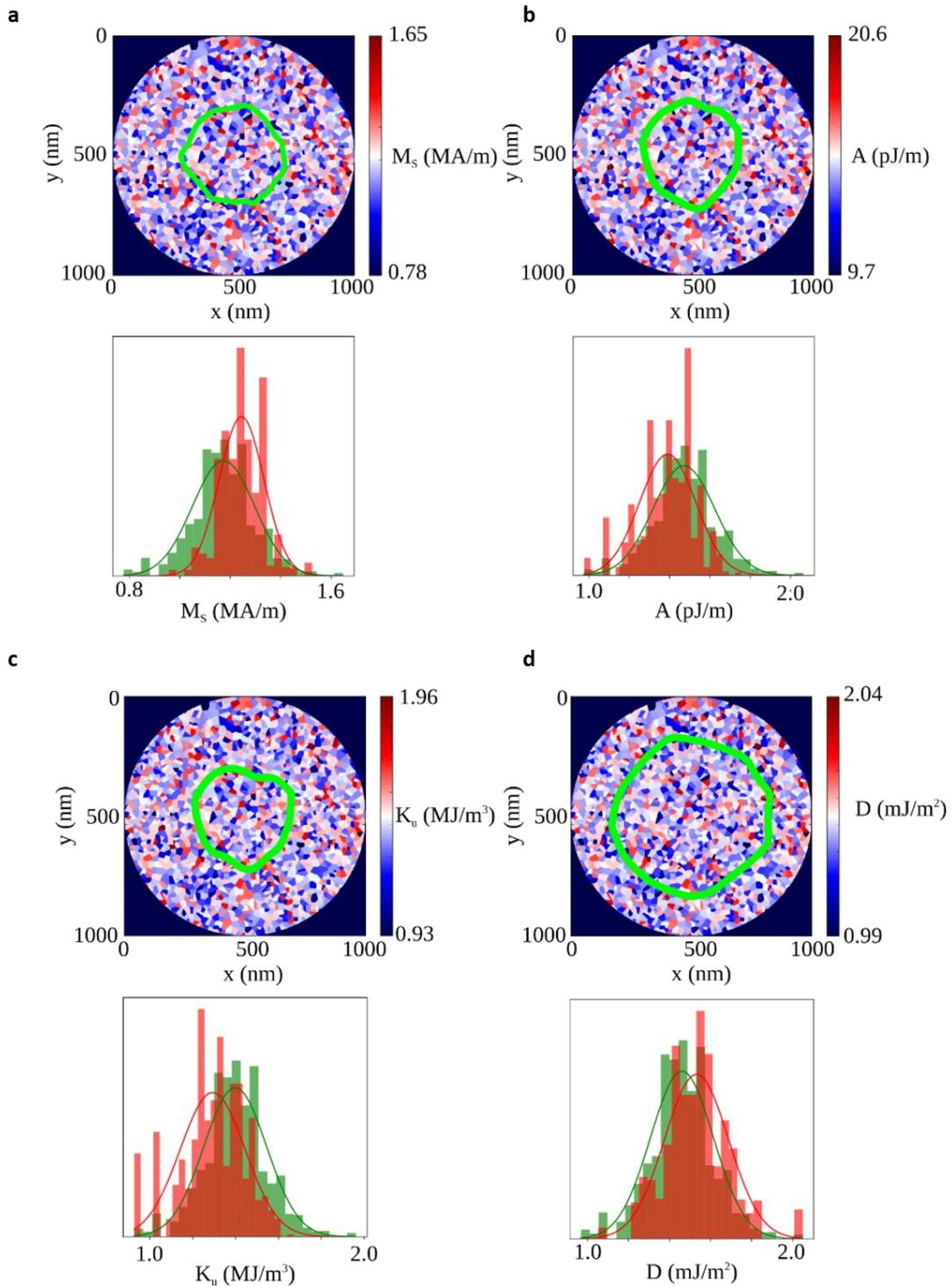

Figure 4: Magnetic parameter disorder in a 7 nm thick Co disc using a 20 nm grain size. (a) Saturation magnetisation distribution with $\delta M_S/M_S$=10%. (b) Exchange stiffness distribution with $\delta A/A$=10 %. (c) Anisotropy constant distribution with $\delta K_u/K_u$=10 %. (d) DMI strength D distribution with with $\delta D/D$=10 %. In all cases the parameter distribution plotted in the histogram over all grains is shown in green and considering only the skyrmion edge grains is plotted in red.

REFERENCES

1. Fert A, Cros V, Sampaio J. Skyrmions on the track. *Nat Nanotechnol* **8**, 152-156 (2013).

2. Fert AR. Magnetic and Transport Properties of Metallic Multilayers. *Materials Science Forum* **59-60**, 439 (1990).

3. Heinze S*, et al.* Spontaneous atomic-scale magnetic skyrmion lattice in two dimensions. *Nat Phys* **7**, 713-718 (2011).

4. Nagaosa N, Tokura Y. Topological properties and dynamics of magnetic skyrmions. *Nat Nanotechnol* **8**, 899-911 (2013).

5. Romming N, Kubetzka A, Hanneken C, von Bergmann K, Wiesendanger R. Field-Dependent Size and Shape of Single Magnetic Skyrmions. *Phys Rev Lett* **114**, 177203 (2015).

6. Jiang WJ*, et al.* Blowing magnetic skyrmion bubbles. *Science* **349**, 283-286 (2015).

7. Jonietz F*, et al.* Spin Transfer Torques in MnSi at Ultralow Current Densities. *Science* **330**, 1648-1651 (2010).

8. Litzius K*, et al.* Skyrmion Hall effect revealed by direct time-resolved X-ray microscopy. *Nat Phys* **13**, 170-175 (2017).

9. Woo S*, et al.* Observation of room-temperature magnetic skyrmions and their current-driven dynamics in ultrathin metallic ferromagnets. *Nat Mater* **15**, 501-+ (2016).

10. Braun HB. Topological effects in nanomagnetism: from superparamagnetism to chiral quantum solitons. *Adv Phys* **61**, 1-116 (2012).

11. Boulle O*, et al.* Room-temperature chiral magnetic skyrmions in ultrathin magnetic nanostructures. *Nat Nanotechnol* **11**, 449-+ (2016).

12. Hrabec A*, et al.* Current-induced skyrmion generation and dynamics in symmetric bilayers. *ArXiv* **1611.00647**, (2016).

13. Moreau-Luchaire C*, et al.* Additive interfacial chiral interaction in multilayers for stabilization of small individual skyrmions at room temperature (vol 11, pg 444, 2016). *Nat Nanotechnol* **11**, 731-731 (2016).

14. Romming N*, et al.* Writing and Deleting Single Magnetic Skyrmions. *Science* **341**, 636-639 (2013).

15. Pulecio JF, Hrabec A, Zeissler K, White RM, Zhu Y, Marrows CH. Hedgehog skyrmion bubbles in ultrathin films with interfacial Dzyaloshinskii-Moriya interactions. *ArXiv* **1611.06869**, (2016).

16. Kim JV, Yoo MW. Current-driven skyrmion dynamics in disordered films. *Appl Phys Lett* **110**, 132404 (2017).

17. Iwasaki J, Mochizuki M, Nagaosa N. Current-induced skyrmion dynamics in constricted geometries. *Nat Nanotechnol* **8**, 742-747 (2013).

18. Kim JV, Yoo MW. Current-driven skyrmion dynamics in disordered films. *ArXiv* **1701.08357**, (2017).

19. Lin SZ, Reichhardt C, Batista CD, Saxena A. Particle model for skyrmions in metallic chiral magnets: Dynamics, pinning, and creep. *Phys Rev B* **87**, 214419 (2013).

20. Muller J, Rosch A. Capturing of a magnetic skyrmion with a hole. *Phys Rev B* **91**, 054410 (2015).

21. Reichhardt C, Ray D, Reichhardt CJO. Collective Transport Properties of Driven Skyrmions with Random Disorder. *Phys Rev Lett* **114**, 217202 (2015).

22. Hanneken C, Kubetzka A, von Bergmann K, Wiesendanger R. Pinning and movement of individual nanoscale magnetic skyrmions via defects. *New J Phys* **18**, (2016).

23. Bogdanov A, Hubert A. The Properties of Isolated Magnetic Vortices. *Phys Status Solidi B* **186**, 527-543 (1994).

24. Wilson MN, Butenko AB, Bogdanov AN, Monchesky TL. Chiral skyrmions in cubic helimagnet films: The role of uniaxial anisotropy. *Phys Rev B* **89**, 094411 (2014).

25. Geissler J*, et al.* Pt magnetization profile in a Pt/Co bilayer studied by resonant magnetic x-ray reflectometry. *Phys Rev B* **65**, 020405 (2002).

26. Gross I*, et al.* Direct measurement of interfacial Dzyaloshinskii-Moriya interaction in X vertical bar CoFeB vertical bar MgO heterostructures with a scanning NV magnetometer (X=Ta, TaN, and W). *Phys Rev B* **94**, 064413 (2016).

27. Soucaille R*, et al.* Probing the Dzyaloshinskii-Moriya interaction in CoFeB ultrathin films using domain wall creep and Brillouin light spectroscopy. *Phys Rev B* **94**, 104431 (2016).

28. Hrabec A*, et al.* Measuring and tailoring the Dzyaloshinskii-Moriya interaction in perpendicularly magnetized thin films. *Phys Rev B* **90**, 020402 (2014).

29. Shepley PM, Tunnicliffe H, Shahbazi K, Burnell G, Moore TA. Tuning domain wall energy with strain: balancing anisotropy and exchange energies in Pt/Co/Ir. *ArXiv* **1703.05749**, (2017).

30. Vansteenkiste A, Leliaert J, Dvornik M, Helsen M, Garcia-Sanchez F, Van Waeyenberge B. The design and verification of MuMax3. *Aip Adv* **4**, (2014).

31. Garcia-Sanchez F, Sampaio J, Reyren N, Cros V, Kim JV. A skyrmion-based spin-torque nano-oscillator. *New J Phys* **18**, (2016).

32. Akbulut S, Akbulut A, Ouml;zdemir M, Yildiz F. Effect of deposition technique of Ni on the perpendicular magnetic anisotropy in Co/Ni multilayers. *J Magn Magn Mater* **390**, 137-141 (2015).

33. Raabe J*, et al.* PolLux: A new facility for soft x-ray spectromicroscopy at the Swiss Light Source. *Rev Sci Instrum* **79**, (2008).

34. Fangohr H, Bordignon G, Franchin M, Knittel A, de Groot PAJ, Fischbacher T. A new approach to (quasi) periodic boundary conditions in micromagnetics: The macrogeometry. *J Appl Phys* **105**, (2009).

35. Fischbacher T, Franchin M, Bordignon G, Fangohr H. A systematic approach to multiphysics extensions of finite-element-based micromagnetic simulations: Nmag. *Ieee T Magn* **43**, 2896-2898 (2007).

36. Aharoni A. *Introduction to Theory of Ferromagnetism*. Oxford University Press (2000).

37. Chikazumi S. *Physics of Ferromagnetism*. Oxford University Press (1997).

38. Je SG, Kim DH, Yoo SC, Min BC, Lee KJ, Choe SB. Asymmetric magnetic domain-wall motion by the Dzyaloshinskii-Moriya interaction. *Phys Rev B* **88**, 214401 (2013).

39. Kostylev M. Interface boundary conditions for dynamic magnetization and spin wave dynamics in a ferromagnetic layer with the interface Dzyaloshinskii-Moriya interaction. *J Appl Phys* **115**, (2014).

40. Moon JH*, et al.* Spin-wave propagation in the presence of interfacial Dzyaloshinskii-Moriya interaction. *Phys Rev B* **88**, 184404 (2013).

**SUPPLEMENTARY INFORMATION**

**Pinning and hysteresis in the field dependent diameter evolution of skyrmions in Pt/Co/Ir superlattice stacks**

EXCHANGE STIFFNESS MEASUREMENTS

The exchange stiffness was extracted from the temperature dependence of the saturation magnetisation which can be described by the Bloch law sufficiently far below the Curie temperature. The Bloch law is given by [1,2]

$$M(T)/M_S = 1 - C\left(\frac{k_B T 4 S^2}{Aa}\right)^{3/2}, \qquad (S1)$$

where $C$ = 0.0294 in case of the fcc lattice that is the case here, $S$ = 1, $k_B$ is the Boltzmann constant, $T$ is the temperature, and $a$ = 0.355 nm is the lattice constant.

DMI MEASUREMENTS

Two methods were used to extract the DMI strength: asymmetric bubble expansion and Brillouin light scattering (BLS). The bubble expansion gives a localized value of $D$ at pinning sites whereas BLS returns the average measured over a larger area and hence is less susceptible to local pinning potentials.

In the asymmetric bubble expansion method, a bubble was nucleated in a single trilayer [3,4]. An out-of-plane magnetic field was pulsed and domain wall creep velocity was extracted from the visible bubble expansion. The domain wall velocity was measured for different static in-plane fields which act with or against the DMI field at the domain wall on either side of the bubble, modifying the wall energy (see figure Supplementary 1 (a)). The minima in the velocity versus in-plane field shows the point where the in plane field cancels the effective in plane field induced at the wall by the DMI, $H_{DMI}$, and can be used to extract the DMI strength $D$ using the formula

$$D = \mu_0 H_{DMI} M_S \Delta, \qquad (S2)$$

where $M_S$ is the saturation magnetisation and $\Delta$ is the domain wall width and is given by $\sqrt{A/K_{eff}}$, in which $A$ is the exchange stiffness and $K_{eff}$ is the effective perpendicular anisotropy constant.

In the Brillouin light scattering method [5,6], the DMI strength is extracted by measuring an asymmetry in the Stokes and anti-Stokes frequencies of light that has been inelastically scattered from propagating spin waves. In a sample with notable DMI strength, spin waves of a given wavelength propagating in opposite directions have different energies. This behaviour is known as propagation nonreciprocity and occurs when the sample is magnetised in-plane and the spin wave vector is perpendicular to the magnetisation, the Damon-Eshbach geometry. The frequency shifts of the inelastically scattered light with respect to the incident laser beam frequency is directly proportional to the DMI strength

$$\Delta f = f_S - f_{AS} = \frac{2\gamma}{\pi M_S} D k_{SW}, \qquad (S3)$$

where $k_{SW}$ is the magnon wavevector, $f_S$ is the Stokes frequency, $f_{AS}$ is the anti-Stokes frequency, and $\gamma$ is the gyromagnetic ratio. DMI was calculated using $D=(\Delta f\ \pi\ M_S\ t)/(2\gamma k)$ where $\gamma$=190 GHz/T is the gyromagnetic ratio, and t is the thickness. The saturation magnetisation $M_S$ was measured to be 1.1 MA/m for this sample. The Stokes and anti-Stokes spectra measured for $k$=7.1 can be seen in supplementary figure 1 (b). Supplementary figure 1 (c) shows the frequency shift for different wave vectors at positive and negative field. The average $D$ was found to be 0.93±0.1 mJ/m².

EFFECTIVE THICKNESS MICROMAGNETIC SIMULATION

Supplementary figure 1 (d) shows the simulated defect free skyrmion diameter as function of magnetization. The *N*=10 repeat stack is compared to the effective thickness simulation where only one thick magnetic layer is considered. This shows that an effective thickness simulation is a good approximation of the multilayer system.

GRAIN SIZE OF POLYCRYSTALINE Pt/Co/Ir

Bright field transmission electron microscopy images where acquired on a representative trilayer Pt/Co/Ir polycrystalline sample. Ten images where taken at x50k and ten images where taken x80k magnification (see supplementary figure 1 (e) for an image taken at x50k). In total 20 different locations on the sample were investigated. Supplementary figure 1 (f) shows the percentage area of the different grain sizes observed, whilst the inset shows the frequency of the observed grains sizes. Whilst around 2/3 of the sample area is covered by grains less than 7 nm diameter, a second peak in the distribution was observed around 10 nm with 1/3 of the sample area being covered with these larger grains. This bi-modal distribution is not atypical for sputtered films in this thickness range. According to Kim et al. [7] magnetic skyrmions get pinned most strongly by grains of comparable size. In our structures the skyrmion has a diameter of 130 nm-270 nm and hence modelling the disorder with the average large grain size of 10 nm is reasonable for the observed distribution of grain sizes.

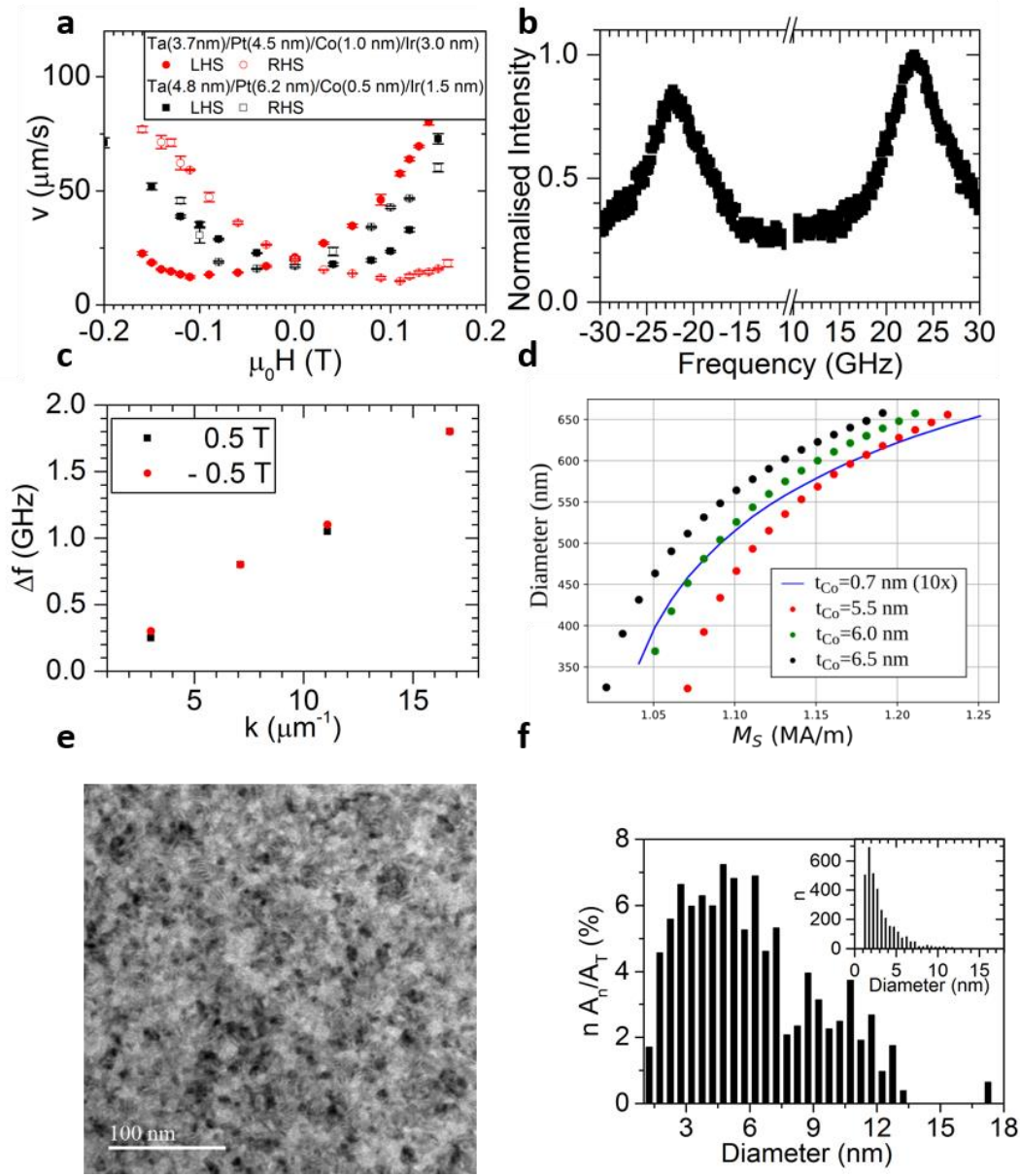

Supplementary Figure 1: (a) Domain wall creep velocity dependence on an in-plane field. The minima shows then DMI field. (b) BLS Stokes and anti-Stokes peak at 0.5 T in plane field. The maximum peak frequency shift was measured to be $\Delta f$=0.8 GHz with an incident angle of 17.5° corresponding to a spin wave vector, $k$ = 7.1 µm$^{-1}$. (c) Frequency shift versus wave vector in the BLS measurement. (d) Skyrmion diameter as function of magnetization simulated for a multilayer stack of 10 Co layers in comparison to simulations of a single Co layers with a similar effective thicknesses. This shows that an effective thickness simulation is a good approximation of the multilayer system. (e) Bright field transmission electron microscopy image of a representative polycrystalline trilayer Pt/Co/Ir sample. (f) The percentage area coverage as a function of grain diameter, by grains of different sizes, $nA_n/A_T$ where $n$ is the number of grains with the same diameter, $A_n$ is the area associated with such a grain and $A_T$ is the total area taken up by the measured grains. The inset shows the frequency of the observed grain sizes.

1. Aharoni A. *Introduction to Theory of Ferromagnetism*. Oxford University Press (2000).

2. Chikazumi S. *Physics of Ferromagnetism*. Oxford University Press (1997).

3. Hrabec A*, et al.* Measuring and tailoring the Dzyaloshinskii-Moriya interaction in perpendicularly magnetized thin films. *Phys Rev B* **90**, 020402 (2014).

4. Je SG, Kim DH, Yoo SC, Min BC, Lee KJ, Choe SB. Asymmetric magnetic domain-wall motion by the Dzyaloshinskii-Moriya interaction. *Phys Rev B* **88**, 214401 (2013).

5. Kostylev M. Interface boundary conditions for dynamic magnetization and spin wave dynamics in a ferromagnetic layer with the interface Dzyaloshinskii-Moriya interaction. *J Appl Phys* **115**,  (2014).

6. Moon JH*, et al.* Spin-wave propagation in the presence of interfacial Dzyaloshinskii-Moriya interaction. *Phys Rev B* **88**, 184404 (2013).

7. Kim JV, Yoo MW. Current-driven skyrmion dynamics in disordered films. *Appl Phys Lett* **110**, 132404 (2017).